\def\selectedoptions{final}

\ifx\empty\selectedoptions
  \def\selectedoptions{final}
\fi

\documentclass[
   \selectedoptions
  ]
  {aipproc}

\def\selectedlayoutstyle {8x11double}
\layoutstyle\selectedlayoutstyle

\SetInternalRegister\hbadness{8000}

\newcommand\doingARLO[2][]{%
  \ifx\mmref\undefined #1\else #2\fi
}

\begin{document}

\def\apj{{\it Ap.\ J.,}}
\def\apjl{{\it Ap.\ J. Letters,}}
\def\aa{{\it Astron.\ Astrophys.,}}
\def\etal{{\it et al.}}
\def\annrev{{\it Ann.\ Rev.\ Astron.\ Ap.}}
\def\aplet{{\it Ap.\ Letters}}
\def\aj{{\it Astron.\ J.}}
\def\apj{Ap J}
\def\apjl{{\it Ap.J. Letters}}
\def\apjlet{{\it Ap.\ J.\ (Lett.)}}
\def\apjs{{\it Ap.\ J.\ Suppl.}}
\def\apjsup{{\it Ap.\ J.\ Suppl.}}
\def\aasup{{\it Astron.\ Astrophys.\ Suppl.}}
\def\aap{{\it Astron.\ Astrophys.\ Suppl.}}
\def\astro{{\it Astron.\ Astrophys.}}
\def\aa{{\it Astron.\ Astrophys.}}
\def\mnras{{\it M.\ N.\ R.\ A.\ S.}}
\def\iaucirc{{\it IAU Circular No .}}
\def\nature{{\it Nature}}
\def\nat{{\it Nature}}
\def\pasa{{\it Proc.\ Astr.\ Soc.\ Aust.}}
\def\pasp{{\it P.\ A.\ S.\ P.}}
\def\pasj{{\it PASJ}}
\def\pre{{\it Preprint}}
\def\qjras{{\it Quart.\ J.\ R.\ A.\ S.}}
\def\rppp{{\it Rep.\ ProAg.\ Phys.}}
\def\sovlet{{\it Sov. Astron. Lett.}}
\def\adspr{{\it Adv. Space. Res.}}
\def\expas{{\it Experimental Astron.}}
\def\ssr{{\it Space Sci. Rev.}}
\def\inpress{in press.}
\def\souspresse{sous presse.}
\def\inprep{in preparation.}
\def\enprep{en pr\'eparation.}
\def\submit{submitted.}
\def\soumis{soumis.}
\def\kte{kT$_{\rm e}$}
\def\ktbb{kT$_{\rm BB}$}
\def\rbb{R$_{\rm BB}$}
\def\ergs{ergs s$^{-1}$}
\def\erg{erg}
\def\ergscm{ergs s$^{-1}$ cm$^{-2}$}
\def\mdot{$\dot{\rm M}$}
\def\rin{R$_{\rm in}$}
\def\ledd{L$_{\rm Edd}$}
\def\comptt{{\sc Comptt}}
\def\msol{M$_\dot$}
\def\nh{N$_{\rm H}$}
\def\meszaros{M\'esz\'aros}
\def\eclairs{{\it ECLAIRs}}
\def\swift{{\it SWIFT}}
\def\glast{{\it GLAST}}
\def\hete{{\it HETE-2}}
\def\egret{{\it EGRET}}
\def\bepposax{{\it Beppo-SAX}}
\def\ginga{{\it GINGA}}
\def\etal{{et al.}}
\def\ga{\hbox{\rlap{\raise.3ex\hbox{$>$}}\lower.8ex\hbox{$\sim$}\ }}

\title 
      [ECLAIRs: A microsatellite to observe the prompt optical and X-ray
emission of Gamma-Ray Bursts]
      {ECLAIRs: A microsatellite to observe the prompt optical and X-ray
emission of Gamma-Ray Bursts}


\author{Didier Barret}{
  address={Centre d'Etude Spatiale des Rayonnements, CNRS/UPS, 31028 Toulouse Cedex 04, France},
  email={didier.barret@cesr.fr},
}
\copyrightyear  {2001}

\begin{abstract} The prompt $\gamma$-ray emission of Gamma-Ray Bursts
(GRBs) is currently interpreted in terms of radiation from electrons
accelerated in internal shocks in a relativistic fireball.  On the
other hand, the origin of the prompt (and early afterglow) optical and
X-ray emission is still debated, mostly because very few data exist
for comparison with theoretical predictions.  It is however commonly
agreed that this emission hides important clues on the GRB physics and
can be used to constrain the fireball parameters, the acceleration and
emission processes and to probe the surroundings of the GRBs. 
\eclairs~is a microsatellite devoted to the observation of the prompt
optical and X-ray emission of GRBs.  For about 100 GRBs yr$^{-1}$,
{\it independent of their duration}, \eclairs~will provide high time
resolution high sensitivity spectral coverage from a few eV up to
$\sim 50$ keV and localization to $\sim 5$'' in near real time.  This
capability is achieved by combining wide field optical and X-ray
cameras sharing a common field of view ($\ga 2.2$ steradians) with the
coded-mask imaging telescopes providing the triggers and the coarse
localizations of the bursts.  Given the delays to start ground-based
observations in response to a GRB trigger, \eclairs~is unique in its
ability to observe the early phases (the first $\sim 20$ sec) of all
GRBs at optical wavelengths.  Furthermore, with its mode of operation,
\eclairs~will enable to search for optical and X-ray precursors
expected from theoretical grounds.  Finally \eclairs~is proposed to
operate simultaneously with \glast~on a synchronous orbit.  This
combination will ensure broad band spectral coverage from eV to GeV energies for the GRBs detected by the two satellites, \eclairs~further providing their
accurate localization to enable follow-up studies. 
\eclairs~relies upon an international collaboration involving
theoretical and hardware groups from Europe and the United States.  In
particular, it builds upon the extensive knowledge and expertise that
is currently being gained with the \hete~mission.
\end{abstract}

\date{\today}

\maketitle

\section{Introduction}
GRBs occur at cosmological distances and are the most violent
explosive phenomena presently observed in the Universe.  For the
strongest GRBs, up to $\sim 10^{54}$ erg (assuming isotropic emission)
can be radiated in the $\gamma$-ray domain on a very short timescale
(from milliseconds to $\sim 100$ seconds).  In the currently favored
models, GRBs are associated with the collapse of massive stars
(collapsar, \cite{woo93,pac98}) or mergers of two compact
stars (two neutron stars or a neutron star and a black hole, e.g.,
\cite{eic89}) (see e.g., \cite{piran99,cas01} for
recent reviews).  Since GRBs are observable across the whole Universe,
if they are indeed linked to the ultimate stages of massive star
evolution, their redshift distribution should reveal the formation
rate of massive stars up to very high redshifts, thus making GRBs
effective cosmological probes (see e.g., \cite{ram00,lam00}).

In both the collapsar and the merger models, the final product is a
stellar mass black hole and a rapidly rotating torus, from which the
energy can be extracted via magnetohydrodynamic processes.  The
energy release in a very small volume produces a relativistic fireball
with a Lorentz factor of at least a few hundred (e.g., \cite{sar97}).  When the relativistic flow decelerates in the interstellar
medium (ISM), a forward and a reverse external shock are produced. 
The forward external shock can account for the afterglow emission
observed at radio, optical and X-ray wavelengths (\cite{djo01} for a recent review).

Whereas the physics of the afterglow is relatively well understood,
the origin of the prompt emission is still debated, especially in
X-rays and optical.  In the relativistic fireball model, the Lorentz
factor of the wind is supposed to be variable so that successive
shells of plasma have large relative velocities leading to the
formation of internal shocks (\cite{ree94,dai00,bel00}).  In that model, the non-thermal
$\gamma$-ray emission is associated with either synchrotron or inverse
Compton emission of electrons accelerated in those shocks.  In X-rays
and optical, the picture is not as clear, mostly because there are not
as many high quality observations available in that energy range
compared to the $\gamma$-ray domain.  There is however growing
observational and theoretical evidence that this energy domain
contains critical information on the GRB physics, the nature of the
progenitors, the way the initial bulk energy is converted into
electromagnetic radiation.  A better understanding of the GRB physics
is required to test models predicting that GRBs may be sources of
ultra high-energy cosmic rays, neutrinos, gravitational waves (see
\cite{pos01} for a recent review).  This understanding is also
required if one intends to make GRBs a reliable tool for cosmology and
for the study of the Universe at very high redshifts.

\eclairs~is a microsatellite proposed to the French Space Agency (CNES). It is specifically devoted to the observation of the prompt optical and X-ray emission of GRBs. 
In the next section, we briefly describe the main scientific objectives of
\eclairs~emphasizing on the area where it will bring an outstanding
contribution.  We then present the science payload and mission concept. 
Finally we emphasize on the complementarity of \eclairs~with two other
missions (\glast~and \swift) supposed to fly simultaneously with
\eclairs.
%
% SCIENCE OBJECTIVES
%
\section{Scientific objectives}
\label{scienceobjectives}
With \eclairs~we wish to use, for the short and long duration GRBs,
the prompt optical and X-ray emission to probe 1) the physics at work
during the event and 2) the surrounding of the burst to get insights
on their origin.  In addition, thanks to its instrumental
capabilities, with \eclairs, we will investigate the existence of
optical and X-ray precursors expected from theoretical grounds, and
whose presence would put unprecedented constraints on any GRB models.

As far as the prompt emission is concerned, as discussed above, very
few data exist in optical and X-rays in contrast to the $\gamma$-ray
domain.  In the optical, so far only one GRB has been detected (by the
ROTSE automated telescope; GRB990123, \cite{ake99}), and for a
few others, upper limits are available for the late part ($\ge 10-20$
second after the onset) of the events (e.g. \cite{boe01}).  In
X-rays, the situation is slightly better, mostly thanks to the
\ginga~(e.g., \cite{mur91}) and \bepposax~satellites
\cite{fro00}.

\subsection{What can be learned from the prompt optical emission?}
It has suggested that the prompt optical emission
as the one observed in GRB990123 may be associated with electrons
accelerated in the reverse external shock \cite{sar99}.  The strength of the
optical emission depends on various parameters, but can in principle
yield constraints on the wind initial Lorentz factor and the
interstellar medium density \cite{sar99,kob00}.
Alternatively, the prompt optical emission could arise from the
forward shock of the blast wave when it propagates in the
pre-accelerated and pair-loaded environment \cite{bel02}.  This
emission can also give constraints on the radiation process itself;
i.e., synchrotron versus Inverse Compton emission; a much stronger
optical flash is expected in the Inverse Compton scenario (e.g.,
\cite{dai98}).

The reasonable question to ask is why the prompt optical emission has
so far been observed from only one GRB (GRB990123, \cite{ake99}).  The poor location accuracy of GRB detectors (e.g., BATSE), the
delays in getting the finalized positions to the ground, the limited
observing efficiency of automated optical telescopes, their response
time, all conspire to make sensitive (below mag $\sim$ 14) and truly
simultaneous observations of GRBs over the whole event almost
impossible.  For ROTSE, the shortest response time that has been
achieved is $\sim 10$ seconds \cite{keh01}.  \eclairs~will not
face any of these problems as the optical and X-ray cameras will
operate continuously over a common field of view.  Optical
coverage will thus be granted for all types of bursts, independently
of their duration, before, during and even after the event.  This
unique capability will also offer the opportunity to study the
transition between the prompt and early afterglow phases at similar
wavelengths.

Extinction by dust in the host galaxy may naturally prevent the
detection of the prompt optical emission.  This is the argument used
to explain the lack of optical emission in some afterglows, otherwise
detected in X-rays and in radio.  Dust extinction is not unexpected if
GRBs are associated with massive star formation.  Djorgovski et al. 
(2001) have however found that the {\it maximum} fraction of optical
afterglows hidden by dust is $\sim 50$\%.  This is an upper limit, as
some optical afterglows may have been missed for various reasons: very
high redshifts, rapid decline rate, intrinsic faintness.

\eclairs~will seek optical emission down to magnitude $\sim 15$ (R
band, 8 sec) for all GRBs.  The properties of the prompt optical
emission will be correlated with the properties of the afterglow
optical emission, thus providing complementary constraints on the
relativistic flow parameters and the surroundings of the event.

\subsection{What can be learned from the prompt X-ray emission?}
Let us now consider the prompt X-ray emission.  The good correlation
between the temporal behaviour of the prompt X-ray and $\gamma$-ray
emissions suggests that it is also produced in internal shocks.  In
X-rays however, there might be additional contributions from the
reverse shock \cite{dai00}, the forward shock
\cite{bel02} or from the photosphere of the fireball
\cite{mes00}.  With its excellent sensitivity,
\eclairs~will observe the prompt X-ray emission
of all GRBs, allowing detailed time-resolved X-ray spectroscopy
to be performed.  These studies will set constraints on the Lorentz
factor of the wind, its baryon loading, the emission mechanism and the
relative contribution of the various shock regions in the overall
emission.

The importance of observing the prompt X-ray emission has recently
been reinforced by the discoveries of X-ray spectral features in
\bepposax~observations: e.g., a transient absorption edge in GRB990705
\cite{ama00} and a transient emission feature in GRB990712
\cite{fro01} (spectral features are also observed in the
X-ray afterglows, e.g., \cite{piro99}).  Well before these discoveries, it
was predicted that effects of photo-electric absorption and Compton
scattering from the circum-burst material should lead to observable
changes in the intrinsic GRB spectrum, with the introduction of
absorption cutoffs and features such as K-edges and emission lines
\cite{mes98}.  In principle, these features can be used to
determine the density and composition of the ISM in the immediate
vicinity of the GRB, the GRB redshift and possibly the nature of the
GRB progenitor.  For instance, the transient absorption edge observed
from GRB990705 was satisfactorily modeled with photo-electric
absorption by a medium with a large iron abundance, which could have
been left there by a supernova event which occurred about 10 years
before the burst \cite{ama00}.  Similarly the transient
emission feature seen in GRB990712 was shown to be consistent with
thermal emission of a baryon-loaded expanding fireball when it becomes
optically thin \cite{fro01}.  The above interpretations are
however made difficult by the limited statistical quality of the data. 
With its improved sensitivity and good time and spectral resolution,
\eclairs~will be able to observe the prompt emission of GRBs over the
whole event and for all types of events.

\subsection{What are the short bursts ?  What about the X-ray
precursors ?} GRBs display a bimodal distribution in durations; the
border is around 2 seconds with about 25\% of the GRBs with durations
less than that value.  This distribution seems to correlate with
spectral hardness; the shortest GRBs have on average harder spectra
\cite{dez96}.  It seems therefore plausible that the two
distributions represent two distinct, although quite similar, physical
phenomena.  Extremely short GRBs may be due to primordial black hole
evaporation, short GRBs to merging neutron stars, and the long ones to
collapsars (see e.g., \cite{piran99}).  So far, due to observational
limitations, afterglows have only been identified for the long
duration GRBs and very little is known about the short GRBs. 
\eclairs~will have the unique capability to observe both short and
long duration GRBs.  These observations will thus provide clues to the
following questions: How does the multi-wavelength prompt emission of
the short GRBs compare with those of the long GRBs?  How does the
prompt emission relate to the afterglow properties?  What is the
redshift distribution of short GRBs?  Answering these questions will
help in assessing whether the short duration GRBs are of different
nature than the long ones.

By its mode of operation, \eclairs~will also enable us to search and
study X-ray and optical precursors.  X-ray precursors have already
been observed (\cite{mur91,fro00}), arising
between 10 and 100 seconds before the main event.  Several models have
been put forward to explain these X-ray precursors (or soft excesses)
(e.g., \cite{pac98,nak00,mes00}), all
making some specific predictions which require further data to be
tested.  What is the spectrum of the X-ray precursor?  How does it
evolves with time?  How do its properties relate to the properties of
the main event?  These are some of the questions which will be
addressed by \eclairs.  In addition, Paczy\'nski (2001) recently
pointed out that there are theoretical reasons to expect strong
optical flashes preceding GRBs (e.g., \cite{bel02}), the detection
of which would put stringent constraints on the range of parameters
for GRB models.

%
% SCIENCE OBJECTIVES
%
\section{The \eclairs~mission concept}
\label{eclairsmission}
The \eclairs~mission concept results from the scientific goals
described above and is optimized under the stringent constraints of a
microsatellite: 50 kilos, 50 Watts and a total volume of $\sim 60$ cm
$\times$ 60 cm $\times$ 30 cm (length, width, height) available for
the science payload.  A technical assessment study of \eclairs~was carried out by CNES in December 2001. This study showed that \eclairs~was feasible as a microsatellite, albeit with not much margins. \eclairs~is now lining up for a selection at the end of 2002. This clearly leaves some time to work on the optimization of the science payload. To help us in this task, an international Science Advisory Committee (SAC) was set up for the mission. The science payload described below accounts for the results from the CNES study and for the advises received from the  SAC. 

\begin{table}[!t]
    \begin{tabular}{lccc}
	\hline
	& E-LAXT & E-SXC & E-WFOC \\
	\hline
	Band pass & $\sim 3$--50/5-150 keV & 0.4--15 keV & 500-700 nm \\
	Number of units & 2 & 6 (3 pairs) & 4 \\
	Offset angle & $\pm 10^{\circ}$ & $-28^{\circ}, 0, +28^{\circ}$ & $\pm 25^{\circ}$ \\
	Mass (kg) & 14 & 11 & 14 \\
	Power (instrument + electronics) (W) & 18 & 8 & 16 \\
	Field of view (one unit, FWZR) & $\sim 120^{\circ}\times120^{\circ}$ & $\sim 53^{\circ}\times 53^{\circ}$ & $\sim 50^{\circ}\times 50^{\circ}$ \\
	Positionning accuracy & $\sim 0.5^{\circ}$ & 5'' & 5'' \\
	Number of GRBs yr$^{-1}$ (total) & $\sim 100$ & $\sim 100$ & ? \\
	Limiting mag. (R) (S/N=8) & \ldots & \ldots & 14.8, 17.4 (8,
	1000 sec) \\
	\hline
	\end{tabular}
\caption{The ECLAIRs science payload consisting of three instruments:
the ECLAIRs Large Area X-ray Telescope (E-LAXT), the ECLAIRs Soft
X-ray Cameras (E-SXC), the ECLAIRs Wide Field Optical Cameras
(E-WFOC). Two options are considered for E-LAXT: Silicon and higher density detectors (e.g. CdZnTe). Si would cover the energy range 3-50 keV whereas CdZnTe would cover from 5 keV to 150 keV.}
\label{table_payload}
\end{table}
\subsection{The science payload}
\begin{figure}[!t]
\centerline{\hspace*{0.1cm}\resizebox{0.525\textwidth}{!}{\includegraphics{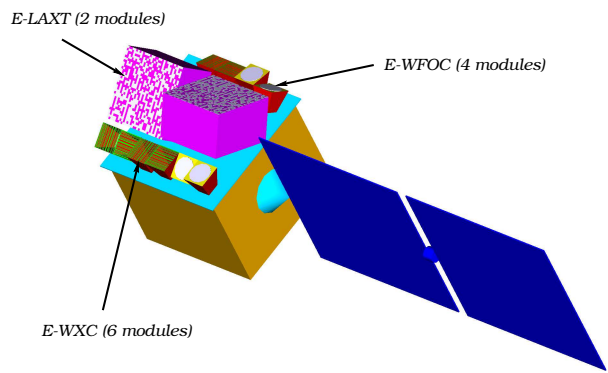}}}
\caption{The \eclairs~science payload on a microsatellite Myriade
spacecraft.  The three sets of instruments which are shown fit within
the geometrical constraints imposed to the science payload.}
\label{eclairs}
\end{figure}

The science payload consists of three sets of instruments (see Table
\ref{table_payload}).  The Large Area X-ray Telescope (E-LAXT), the
Soft X-ray Cameras (E-SXC), and the Wide Field Optical Cameras
(E-WFOC) (see Fig \ref{eclairs}).  It will be provided by a
consortium of institutes which have developed a
considerable expertise along the preparation and operation
of missions, such as \hete~and INTEGRAL. The US contribution to \eclairs~will be the subject of a SMEX/MOO proposal to NASA in 2002.
\subsubsection{ECLAIRs - Large Area X-ray Telescope}
The Large Area X-ray Telescope (E-LAXT) is made of two identical conventional 2D coded-mask imaging telescopes, with offset looking directions. The mask is located 15 cm above the detector. The detector is a pixel semiconductor detector. In the baseline, each pixel was a 2 mm thick Si PIN diodes of 1cm$^2$. The mask cells match the pixel size. Large area Si PIN detectors with their associated low-noise low-power front-end electronics are currently developed at CESR as part of an R\&T program funded by CNES. At low power, the expected noise level should result in an energy resolution of $\sim 1$ keV (at 6 keV, -40C) making possible a low energy threshold of $\sim 3$ keV. The thickness of the diodes ensures an energy coverage up to $\sim 40$ keV: matrix of Si PIN diodes are therefore a possible detector solution for \eclairs.

However, as suggested by the SAC, there are alternatives to Silicon for the E-LAXT detector: CdTe, CdZnTe. Both would extend the energy range of \eclairs~in the hard X-ray range, which would help for the detection of GRBs. CdZnTe, as the ones developed for \swift~or AXO \cite{bud01} have excellent performances for a mW/cm$^2$ ratio similar to Silicon. In addition, using strip readout techniques they have been demonstrated to work down to 5 keV (e.g. \cite{bud01}). Considering CdZnTe, the detector of one E-LAXT could have an effective area of $25 \times  25$ cm$^2$, covered with pixels of $5 \times  5$ mm$^2$ and 2 mm thickness. The imaging system would have an angular resolution of 2 degrees and a positioning accuracy of $\sim 0.5$ degree. Due to the stringent mass constraints on a microsatellite, the mask and shielding will be effective only below  $\sim 50$ keV. Whereas the trigger will be obtained at energies above $\sim$ 50 keV, the position of the GRB will be derived from the images reconstructed below that energy. Using a simplified model for the E-LAXT and the Log(N)/Log(P) curve derived from the BATSE 4B catalog \cite{paciesas99}, we have estimated the rate of GRBs localized by the two E-LAXT units to be larger than $100$ GRB yr$^{-1}$.

\subsubsection{ECLAIRs - Soft X-ray Cameras} The Soft X-ray Cameras
(E-SXC) for \eclairs~are based upon the successfully-flown \hete~design \cite{ric01} (see also these proceedings).  The operating principle is that of a coded-mask
imager, in which a 1-D coded mask is rigidly suspended above an X-ray
charge-coupled device (CCDID-34).  The E-SXC assembly is made of 6
camera modules, covering a field of view of 2.7 sr.  The CCDID-34
(3K$\times$6K array; 10$\mu$m square pixels, 20'') has an overall size
of 30 mm$\times$ 60 mm and is currently in production at MIT Lincoln
Laboratory.  It improves over the CCID-20 used for \hete~by a greater
energy coverage (0.4-15 keV versus 0.8-10 keV), a better time
resolution (0.25 sec versus 1 sec), and a better quantum efficiency
(sensitivity of $\sim 400$ mCrab, 1 sec, $4 \sigma$).  The E-SXC will
provide ~5'' burst localizations (at S/N =8).  About 100 GRBs
yr$^{-1}$ should be detected in the 6 units.
\subsubsection{ECLAIRs - Wide-Field Optical Cameras}
The Wide-Field Optical Cameras (E-WFOC) for \eclairs~are derived from
the star camera units successfully flown on \hete~\cite{ric01}.  The
large field of view  is achieved by four such cameras.  The
limiting magnitude in R is 14.8 (8 s at S/N=8) for one E-WFOC. Each of
the four modules utilizes a moderately fast, well-corrected optical
lens (focal length of 80 mm, f/0.9) coupled to a 2$\times$2 array of
MIT CCID-34 sensors, resulting in a hybrid focal plane with 6K$\times$
6K pixels; each pixel is 10$\mu$m x 10$\mu$m (25.8'').  The
integration time is 2 seconds.

To achieve the light weight and low power required for \eclairs, the
drive and readout electronics, as well as the digital frame buffer
memory, for the E-SXC and E-WFOC instruments will be combined to the
maximum degree possible.

The operating mode for the E-WFOC relies upon digitizing and storing
successive 300 MB image frames in a four stage deep buffer, requiring
a total of 1.2 GB SRAM. In response to triggers from the E-LAXT or
E-SXC, we will select 4$^\circ \times 4^\circ$ regions-of-interest (ie
512 x 512 subarrays) from this large buffer, centered on the suspected
burst coarse localization, for transfer into an optical burst memory. 
In addition, neighborhoods of twenty-five stars, extending out to
64$\times$64 pixels (27'$\times$27'), will also be stored as
astrometric and photometric references.  The accumulation of 500
frames (=1000 sec), each with burst and reference star data, will
reside in 377MB of SRAM, and require 3 minutes to downlink during an
X-band contact with an \eclairs~ground station.  Shift-and-add
summation of the digitized, two-second resolution CCD data in ground
processing will permit the E-WFOC to achieve an ultimate limiting
sensitivity of R=17.4 (1000s at S/N=8).  Centroiding will result in
bright optical transient localizations accurate to $\pm$ 2'', even in
the presence of spacecraft pointing drift (assumed to be 47''/s,
$3\sigma$, as specified for the Myriade spacecraft).  For long term
optical monitoring, we will also be able to downlink 45 full image
frames per day (whole field of view at full angular resolution, every $\sim 30$
minutes), each containing more than 2.5 million star images. 
Downlinking of the full frame data will require 13 GB/day.

We are currently investigating near-infrared cameras (NIRC, 1-2 microns) as an alternative to the E-WFOC. There are three main advantages of considering NIRC. First, in NIR, the extinction is smaller than in the optical. This means that GRBs produced in dusty regions and not visible in the optical might become detectable in NIR. Second, NIRC would have the potential to observe higher z events, due to the Lyman alpha break. Finally, the discovery space is much larger in NIR than in the optical, as the NIR sky has never been searched for variability so far. 

A detailed study is now required to determine whether cooled NIRC with good sensitivity, large field of view can be accommodated on \eclairs~under the stringent constraints on power, mass, etc. of a microsatellite.

\subsection{Implementation of the mission}
The baseline for \eclairs~is an equatorial 550 km (low inclination) orbit for a low, stable
background, low radiation damage to the CCD, and for the download of
the science data to be possible with a single ground station.  This
orbit could be achieved by various launchers, as for example a PEGASUS
from the Marshall Island or from Alcantara.  We plan to reuse the
\hete~ground segment.  In particular, one of the 3 S band stations
(Singapore, Kwajalein, or Cayenne) will be converted to X band.  For
the alert system, we plan to use the \hete~network of 12 VHF stations
located along the equator.  As will be discussed below, \eclairs~is proposed to fly 
simultaneously with \glast; therefore an ideal launch date would be around the end of
2006. The lifetime of the mission is foreseen for 5 years for a
maximum synergy with \glast.

\subsection{Operational considerations}
As far as the attitude control is concerned, the instruments will
point in the anti-earth direction during night time, and look at the pole during day time.  This way the operating temperature of the instruments will be kept low (below -50 C). The triggering system of
\eclairs~is relatively simple.  The on-board computer monitors
continuously the count rates in E-LAXT. When a transient event is
detected, a signal is sent to the E-SXC and E-WFOC for the most recent
data to be stored in a dedicated memory.  Two images from E-LAXT are
then reconstructed before and during the event.  From the difference
of the two images, the rough position ($\sim 0.5$ deg accuracy) of the
event is obtained and sent out to the ground and to the secondary
instruments which use this position to obtain the final more accurate
position.  After $\sim 30$ seconds, the final position (5'' accuracy)
is transmitted to the ground.  During the next passage to the ground
station a high rate X-band communication (16 Mbits/s) allows the whole
data set associated with the event to be downloaded.

The mission and science operations will be performed with the help of CNES control center in Toulouse. The mission and operation center which may be combined will be responsible for receiving the data from CNES, generating the spacecraft commands on a weekly basis, monitoring the health of the spacecraft and science payload, recovering the attitude, and for the quick-look analysis of the data for rapid distribution of the GRB final positions. In addition it will be responsible for the data archives and the education and public outreach
program. The center will likely be provided by a consortium  of institutes including the Geneva observatory, the Strasbourg observatory and the Leicester University X-ray group.

%
% COMPARISON
%
\section{A mission complementary to \glast~and \swift}
\label{glastswift}

\begin{figure}[!t]
\centerline{\hspace*{-0.3cm}\resizebox{0.55\textwidth}{!}{\includegraphics{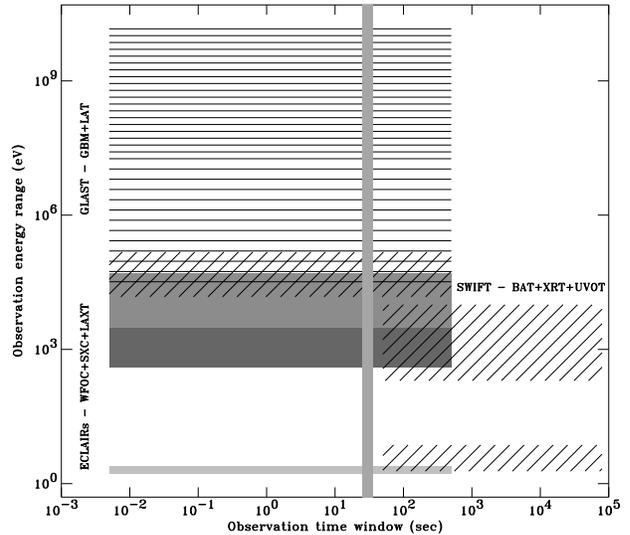}}}
\caption{Comparing \eclairs~(filled regions) with \glast~(horizontal
lines) and \swift~(tilted lines).  The time window of the observations
is given on the X axis whereas the Y axis represents the energy range
of the instruments.  As can be seen, the combination of \eclairs~and
\glast~would provide spectral coverage over 11 decades in energy.  Note
also the complementarity between \eclairs~and \swift; \eclairs~is
focused on the prompt optical/X-ray emission whereas \swift~is designed
for the afterglow emission in the same energy range.  The mean GRB
duration is also shown for indication (vertical box).}
\label{eclairs-swift-glast}
\end{figure}

\glast\footnote{http://www-glast.stanford.edu/}~is scheduled to be
launched in March 2006 into a low earth orbit.  The satellite will
carry 2 instruments: the Large Area Telescope (LAT), which will
observe emission from 20 MeV to 200 GeV, and the Gamma-ray Burst
Monitor (GBM), which will detect transients from 20 keV to 20 MeV. The
LAT detector is ~50 times more sensitive than it's predecessor, \egret.
While only a few GRBs were detected by \egret, \glast~is expected to
observe nearly 200 GRBs per year.  These bursts will also be detected
by the GBM, so that the spectrum will be measured over 7 orders of
magnitude.  Unfortunately, only for the brightest bursts the positions
derived by the LAT will be accurate enough to be used for follow up
observations.  Provided that \eclairs~and \glast~can remain on a
similar orbit (adjustment and maintenance of the orbit can indeed be achieved through the chemical propulsion system of the microsatellite) \eclairs~would greatly enhance the
\glast~science, by both extending the spectra to lower energies (down
to $\sim 2$ eV) and by improving the localizations in near real-time
to enable follow-up observations in the afterglow regime.  Given that
the GBM has a FOV much larger than \eclairs, all GRBs seen by
\eclairs~will be also detected by the GBM, thus providing for $\sim
100$ GRB yr$^{-1}$, spectral coverage from about 7 decades in energy,
and for those detected by the LAT ($\sim 80$) over 11 decades in
energy!  This broad band spectral coverage will enable discrimination
between the various radiation processes proposed for the
multi-wavelength GRB emission, including those, yet to be tested, put
forward for the GeV emission (inverse Compton, Synchrotron emission,
see e.g., \cite{wax97,bot98}).  This will open a
completely new window on the GRB physics, setting for the first time
real constraints on models predicting that GRBs are sources of
ultra-high energy cosmic rays and neutrinos.  Furthermore, the ability
to locate the \glast-GRBs precisely, making possible the identification
of the host galaxies and the measure of the redshifts will enable the
systematics of the GeV emission to be studied, and the \glast-GRBs to
be compared, as a class of events, to the GRBs detected by satellites
operating at lower energies (\bepposax, \hete~and \swift).  Finally,
for those GRBs for which the redshift will be determined, cut-offs in
the observed GeV spectrum can be used to infer the level of
ultra-violet to infra-red background light which is a direct tracer of
star and Galaxy formation in the early Universe \cite{sal98}.  The complementarity between \glast~and \eclairs~is best
illustrated in Fig.  \ref{eclairs-swift-glast} where the observing
energy range is plotted against the observing time window of the
events.  \eclairs~was presented at the last \glast~GRB working group
and received strong support.

\swift\footnote{http://swift.sonoma.edu/}~is a NASA mission dedicated
to the study of the GRB afterglows.  It should be launched in Fall
2003, with a nominal on-orbit lifetime of 3 years.  It will carry
three instruments: The Burst Alert Telescope (BAT) covering the 10 to
150 keV range, and two Narrow Fields Instruments (NFIs); the X-Ray
Telescope (XRT, 0.2-10 keV) and the UV Optical Telescope (UVOT,
170-650 nm).  The observing strategy of \swift~is to point the NFIs
after the detection of a GRB in the BAT. This strategy clearly means
that \swift~will miss the early X-ray and optical emission of all GRBs. 
The time to point the NFIs to the direction of the GRB should range
between 20 and 70 seconds, with a mean value of 50 sec.  Its ability
to observe the precursors and activity during the burst will thus make
\eclairs~a very complementary mission to \swift~(see Fig. 
\ref{eclairs-swift-glast}).
\section{Conclusions}
Fortunately GRBs are extremely bright events which can easily be
detected and studied with an instrumentation matching the stringent
mass and power constraints of a microsatellite.  GRBs have been proved to be highly
complex phenomena whose understanding requires multi-wavelength
observations of the prompt and afterglow phases and follow-up
ground-based observations to determine their host galaxies and their
redshifts.  \eclairs~will thus bring a significant contribution to a
better understanding of GRBs by providing high sensitivity
observations of the prompt optical/X-ray emission and accurate
localization of more than 100 gamma-ray burst per year.
\section{Acknowledgments}
It is a real pleasure for me to thank all the members of the \eclairs~collaboration for their interest in the mission and their support: J.L. Atteia, M. Bo\"er, A. Beloborodov, A. Castro-Tirado, T. Courvoisier, F. Daigne, J.P. Dezalay, B. Dingus, M. Ehanno, P. Goldoni, P. Guillout, J.M. Hameury, G. Henri, K. Hurley, P. Jean, G. Jernigan, J.P. Kneib, D. Lamb, P. Mandrou, A. Marcowith, F. Martel, L. Michel, R. Mochkovitch, C. Motch, J.P. Osborne, M. Pakull, G. Pelletier, J. Poutanen, V. Reglero, G. Ricker, J. Rodrigo, A. Short, R. Svensson and M. Ward. 

A special thank also to C. Meegan and N. Gehrels for their strong support on behalf of the GLAST collaboration.

I also wish to thank the members of  the \eclairs~Science Advisory Committee for their very valuable inputs which will help in optimizing the mission: J.L.  Atteia, A. Castro-Tirado, F. Frontera, J. Hjorth, N. Kawai, S. Kulkarni, C. Meegan, R. Mochkovitch, T. Piran, G. Ricker, B. Stern, S. Woosley. 

I am very grateful to G.K. Skinner, J.L. Atteia and G. Ricker for their continuous help and support.

%%-----------------------------
%%      your bibliography
%%-----------------------------

\end{document}